%% file: IEEE-TPS/ieee.tex
\documentclass[conference]{IEEEtran}

\makeatletter
\def\ps@headings{%
\def\@oddhead{\mbox{}\scriptsize\rightmark \hfil \thepage}%
\def\@evenhead{\scriptsize\thepage \hfil \leftmark\mbox{}}%
\def\@oddfoot{}%
\def\@evenfoot{}}
\makeatother
\IEEEoverridecommandlockouts
\usepackage[letterpaper, left=0.635in,right=0.635in,top=0.75in,bottom=1.04in]{geometry}

\usepackage{cite}
\usepackage{amsmath,amssymb,amsfonts}
\usepackage{algorithmic}
\usepackage{graphicx}
\usepackage{textcomp}
\usepackage{xcolor}
\usepackage{listings}
\usepackage{lipsum}
\usepackage{multirow}
\usepackage{pifont}
\usepackage{newunicodechar}
\usepackage{enumitem}
\usepackage{hyperref}
\usepackage{stfloats}
\usepackage{booktabs}

\def\BibTeX{{\rm B\kern-.05em{\sc i\kern-.025em b}\kern-.08em
    T\kern-.1667em\lower.7ex\hbox{E}\kern-.125emX}}
\begin{document}

\title{ORCAGen: Orchestrating Context-Aware Malware Deception with RAG-Guided Generative AI}

\author{\IEEEauthorblockN{Shihab Ahmed\IEEEauthorrefmark{1}, Md Sajidul Islam Sajid\IEEEauthorrefmark{1}, Teryl Taylor\IEEEauthorrefmark{2}, Frederico Araujo\IEEEauthorrefmark{2}, Tariqul Islam\IEEEauthorrefmark{3}}
	\IEEEauthorblockA{
    \IEEEauthorrefmark{1}Towson University, Towson, MD, USA; 
	\IEEEauthorrefmark{2}IBM Research, Yorktown Heights, NY, USA;\\
    \IEEEauthorrefmark{3}University of Maryland Baltimore County, MD, USA\\
	sahmed23@students.towson.edu, msajid@towson.edu, \{terylt, frederico.araujo\}@ibm.com, mtislam@umbc.edu}
}


\maketitle

\begin{abstract}
Malware defenses often remove or isolate suspicious programs as quickly as possible. While effective for containment, this approach can also waste an opportunity to observe attacker behavior and deploy targeted countermeasures. ORCAGen takes a different approach: it uses GenAI to build malware-specific deception playbooks offline, validates them before deployment, and enforces only the verified logic at runtime. ORCAGen combines Retrieval-Augmented Generation (RAG) with structured prompt engineering to generate both proof-of-concept (PoC) malware and corresponding deception orchestration code. A curated knowledge base (KB) of malware procedures and active defense strategies grounds the generation process, helping the LLM produce threat-specific and executable deception logic rather than generic or hallucinated outputs. The generated PoC malware provides a safe and reproducible way to test whether a deception strategy can disrupt, redirect, or suppress targeted malware behavior before the strategy is added to the runtime playbook. We evaluate ORCAGen across GPT-4o, GPT-5.5, Gemini 3.5 Flash, Qwen3-Coder, and Claude Sonnet 4.5 using execution success, hallucination rate, refinement effort, deception effectiveness, and runtime overhead. The evaluation covers synthesized malware scenarios and 150 real-world malware samples across keyloggers, information stealers, and ransomware. Across the evaluated scenarios, GPT-5.5 required the fewest refinements and produced no observed hallucinated APIs, while Gemini 3.5 Flash achieved the lowest response time and runtime overhead. The results show that RAG-guided structured prompting can support the scalable construction of malware-specific deception playbooks that remain lightweight and deterministic during runtime enforcement.

\end{abstract}

\begin{IEEEkeywords}
Cyber Deception, Generative AI, RAG, LLM, Malware
\end{IEEEkeywords}

\input{IEEE-TPS/file-intro}
\input{IEEE-TPS/file-background}
\input{IEEE-TPS/file-methodology}

\input{IEEE-TPS/file-eval}
\input{IEEE-TPS/file-related-work}
\input{IEEE-TPS/file-conclusion}

\bibliographystyle{IEEEtran}
\bibliography{bibliography}

\clearpage
\appendices

\section{Representative Knowledge Base Entries}\label{app:kb_example}

Table~\ref{tab:kb_example} presents representative entries from the ORCAGen knowledge base. Each entry includes the malware family, attack technique, behavioral description, and corresponding deception strategy.

\begin{table*}
\centering
\renewcommand{\arraystretch}{1.3} 
\setlength{\tabcolsep}{4pt} 
\begin{tabular}{|p{1.4cm}|p{1.3cm}|p{6.2cm}|p{8.1cm}|}
\hline
\textbf{Malware Family} & \textbf{Attack Technique} & \textbf{Behavioral Description} & \textbf{Deception Strategy} \\ \hline

Information Stealer & File System Scanning & Recursively scans directories and locates sensitive files based on target extensions and filenames (e.g., .json, .conf, .ini, .txt, api\_key, browser profiles, wallet.dat). Uses Windows APIs such as \texttt{FindFirstFile} and \texttt{FindNextFile} to scan files in directories (e.g., \verb|C:\Users\*\Documents|) and return the filenames. & \textbf{Filesystem API Hooking:} Intercept calls to file system scanning and access APIs such as \texttt{FindFirstFile, FindNextFile, CreateFile, ReadFile}, etc. Return decoy filenames instead of the actual search results to mislead malware. \\ \hline

Keylogger & Screen Logging & Periodically takes screenshots or captures the display screen when certain events occur. Screen capture APIs (e.g., \texttt{BitBlt, PrintWindow}) capture screen content. & \textbf{Screen Capture Prevention:} The screen capture APIs (e.g., \texttt{BitBlt, PrintWindow}) capture screen content. Intercept screenshot capture APIs and replace captured images with decoy screenshots. For example, hook the \texttt{BitBlt} and \texttt{PrintWindow} APIs to inject decoy images. Ensure legitimate screen capture requests remain functional. \\ \hline

Ransomware & File-Based Encryption & Targets specific file extensions (e.g., .docx, .xlsx, .db, .txt) while skipping critical operating system and application files to keep the system functional. Uses filters or hardcoded lists of extensions to select files for encryption in a specific directory (\verb|C:\Users\*\Documents|). Employs Windows APIs like \texttt{FindFirstFile} and \texttt{FindNextFile} to enumerate files, and \texttt{CreateFile, ReadFile, WriteFile} to read and overwrite file contents with encrypted data. & \textbf{File I/O Hooking:} Intercept file operation APIs such as \texttt{CreateFile, WriteFile, SetFilePointer, FlushFileBuffers} to detect or modify ransomware behavior. Check for suspicious access to large numbers of documents. The goal is to implement a FakeFailure/FakeSuccess approach for specific file types where the ransomware receives error messages, indicating a failure to encrypt files (FakeFailure), or the ransomware perceives the encryption as complete and receives a fake success message while no actual file is encrypted (FakeSuccess), while legitimate file operations remain functional. \\ \hline

\end{tabular}
\vspace{2mm}
\caption{Representative Knowledge Base Entries}
\label{tab:kb_example}
\end{table*}


\section{Knowledge Base Source Distribution}\label{app:KB_distribution}

\begin{table}[!h]
\centering
\setlength{\tabcolsep}{4pt}
\renewcommand{\arraystretch}{1.3}
\begin{tabular}{|p{1.9cm}|p{1.4cm}|p{1.4cm}|p{1.6cm}|p{0.8cm}|} \hline

\textbf{Category} & \textbf{Keylogger} & \textbf{Information Stealer} & \textbf{Ransomware} & \textbf{Total} \\ \hline

Total Documents Collected & 72 & 49 & 35 & 156 \\ \hline

\multicolumn{5}{|c|}{\textit{Retained Documents After Screening}} \\ \hline

Academic Articles & 8 & 9 & 6 & 23 \\ \hline
Threat Reports & 4 & 2 & 2 & 8 \\ \hline
Malware Blogs & 13 & 7 & 8 & 28 \\ \hline
Technical Documentation & 10 & 1 & 1 & 12 \\ \hline

\textbf{Total Unique Documents} & \textbf{35} & \textbf{19} & \textbf{17} & \textbf{71} \\ \hline

\end{tabular}
\vspace{2mm}
\caption{Distribution of Collected and Retained Documents}
\label{tab:kb_distribution}
\end{table}

Table~\ref{tab:kb_distribution} summarizes the document collection and screening process. Candidate sources were collected from multiple categories, including academic articles, threat intelligence reports, malware-analysis blogs, and technical documentation. Academic articles were collected from renowned venues including IEEE, ACM, USENIX, Springer, etc., while industry-oriented knowledge was collected from Mandiant, Malwarebytes, SANS, Microsoft, Android, and MITRE technical documentation, etc. Duplicate sources were removed and documents were screened for technical relevance to malware behaviors or API usage patterns. The retained corpus was then used for structured knowledge extraction.

\end{document}

%% file: IEEE-TPS/file-intro.tex

\section{Introduction}
Malware defense often ends with containment: detect the process, isolate it, and remove it as quickly as possible. While this response is necessary in many cases, it also limits what defenders can learn after compromise. Once malware is terminated, defenders may lose visibility into how it searches for credentials, captures user input, scans files, selects encryption targets, or adapts to the host environment. This creates a practical gap. Modern malware is difficult to detect early because it uses polymorphism, code obfuscation, dynamic API resolution, and anti-analysis techniques, yet immediate removal can prevent defenders from observing the behavior needed to build targeted responses.

Cyber deception offers a complementary approach. Instead of only blocking or removing malicious activity, deception can engage malware in controlled ways by presenting honeyfiles, false credentials, decoy services, or manipulated API responses \cite{virtual_honeypot_framework, virtual_honeypot_botnet_tracking, honeyproxy_design_implementation}. This allows defenders to slow, mislead, or suppress malicious behavior while preserving visibility into the attack. However, most deception systems still depend on manually configured rules and static ploys \cite{anagnostakis2005detecting, urias2017technologies, chimera_autonomous_planning}. As malware behavior changes, these manually engineered strategies become difficult to scale. A practical deception system must therefore construct threat-specific responses, validate them before deployment, and enforce them at runtime without adding fragile decision logic to the attack path.

Generative AI (GenAI) creates a new opportunity for building such deceptive systems. Large language models (LLMs) can generate operational code when guided by structured prompts and domain context, and recent work has shown their usefulness in malware emulation, attack scripting, and controlled security testing \cite{bao2025generating, iturbe2024unleashing}. Yet most GenAI-based deception work focuses on generating static artifacts such as honeytokens, baits, fake user profiles, or interactive responses from general prompts \cite{act_as_honeytoken, chatbots_honeypot_world, engaging_attackers_highly, llm_in_shell}. These systems do not fully address how to generate malware-specific deception logic, test whether it works, and deploy it safely against runtime malware behaviors. Using live LLM inference during malware execution would also introduce latency, reliability, and safety concerns. The key challenge is therefore not simply to generate deception artifacts, but to generate deception logic that can be grounded, tested, and compiled into a reliable runtime playbook.

This paper presents ORCAGen, a GenAI-enabled framework for constructing malware-specific deception playbooks that defenders can deploy after a suspicious process, malware family, or target behavior has been identified. ORCAGen separates playbook construction from runtime enforcement. During the offline phase, it uses Retrieval-Augmented Generation (RAG) and structured prompt engineering to generate both proof-of-concept (PoC) malware and corresponding deception orchestration code. PoC malware provides a safe and reproducible way to emulate behaviors such as keylogging, credential theft, and ransomware-style file operations. The generated deception logic is then compiled, executed, and validated against the PoC behavior. Only verified logic is added to a deployable playbook, which performs runtime enforcement without live LLM inference. During incident response, defenders can inject the validated Super DLL into a suspicious process to disrupt, redirect, or suppress malware behavior through pre-tested interventions instead of immediately terminating the process and losing visibility into attacker actions. At the core of ORCAGen is a curated knowledge base that maps malware procedures to active defense strategies. The knowledge base is built from academic publications, threat intelligence reports, malware blogs, and technical documentation. During generation, RAG retrieves relevant malware procedures and defense strategies and injects them into structured prompt templates. This grounding helps the model generate code that is specific to the target malware behavior rather than generic or hallucinated. To the best of our knowledge, ORCAGen is the first framework to combine GenAI and RAG for offline generation, validation, and compilation of malware-specific deception playbooks that can be enforced at runtime through validated, deterministic logic. The key contributions of this work are as follows:

\begin{itemize}[leftmargin=1em, noitemsep, topsep=1pt]
    \item We construct a structured knowledge base (KB) that maps malware techniques to corresponding active defense strategies, providing the grounding needed for threat-specific deception generation.

    \item We design a RAG-guided prompt architecture that retrieves relevant malware and defense knowledge during offline generation, enabling LLMs to produce behaviorally consistent PoC malware and deception orchestration code without additional model training.

    \item We design an offline playbook generation and validation workflow that adds only verified ploys to the runtime playbook by (i) creating PoC malware, (ii) generating corresponding deception logic, and (iii) testing the deception in a controlled environment.
    
    \item We comprehensively evaluate ORCAGen across multiple LLMs using execution success, hallucination rate, refinement effort, deception effectiveness, and runtime overhead.
\end{itemize}


The remainder of this paper is organized as follows. Section~\ref{sec:background} provides background on GenAI, RAG, and deception-oriented orchestration. Section~\ref{sec:threat_model} provides the threat model. Section~\ref{sec:methodology} presents the ORCAGen design. Section~\ref{sec:evaluation} presents the evaluation. Section~\ref{sec:rw} reviews related work. Section~\ref{sec:conclusion} concludes the paper.

%% file: IEEE-TPS/file-background.tex
\section{Background}\label{sec:background}
\subsection{Generative AI in Security Contexts}
Large Language Models (LLMs) are increasingly used in cybersecurity tasks that require reasoning over technical context and producing structured outputs. Prior work has used LLMs for malware emulation and synthetic malware generation~\cite{bao2025generating}, automated attack-technique code generation~\cite{iturbe2024unleashing}, vulnerability analysis~\cite{yang2026liva}, reverse engineering support~\cite{Hu2024DeGPTOD}, threat-intelligence extraction~\cite{psarrou2026enhanced}, and deception-oriented ploys~\cite{ahmed2025spade}. These applications show that LLMs can translate security context into deployable artifacts that support testing, analysis, and response design. However, practical security use requires more control than a general prompt can provide. LLMs may produce incomplete code, rely on outdated knowledge, or hallucinate APIs and behaviors that do not exist in the target environment \cite{liu2026beyond}. Three common mechanisms are used to reduce these risks. Structured prompt engineering defines the task, output format, constraints, and examples. Retrieval-Augmented Generation (RAG) supplies task-specific external knowledge at generation time. Fine-tuning adapts a model using a domain-specific corpus, but requires additional training and model access. ORCAGen uses RAG with structured prompting rather than fine-tuning, allowing the model to remain unchanged while grounding each generation task in malware procedures and active defense strategies.

\begin{figure*}[!ht]
    \centering
    \includegraphics[width=0.98\textwidth]{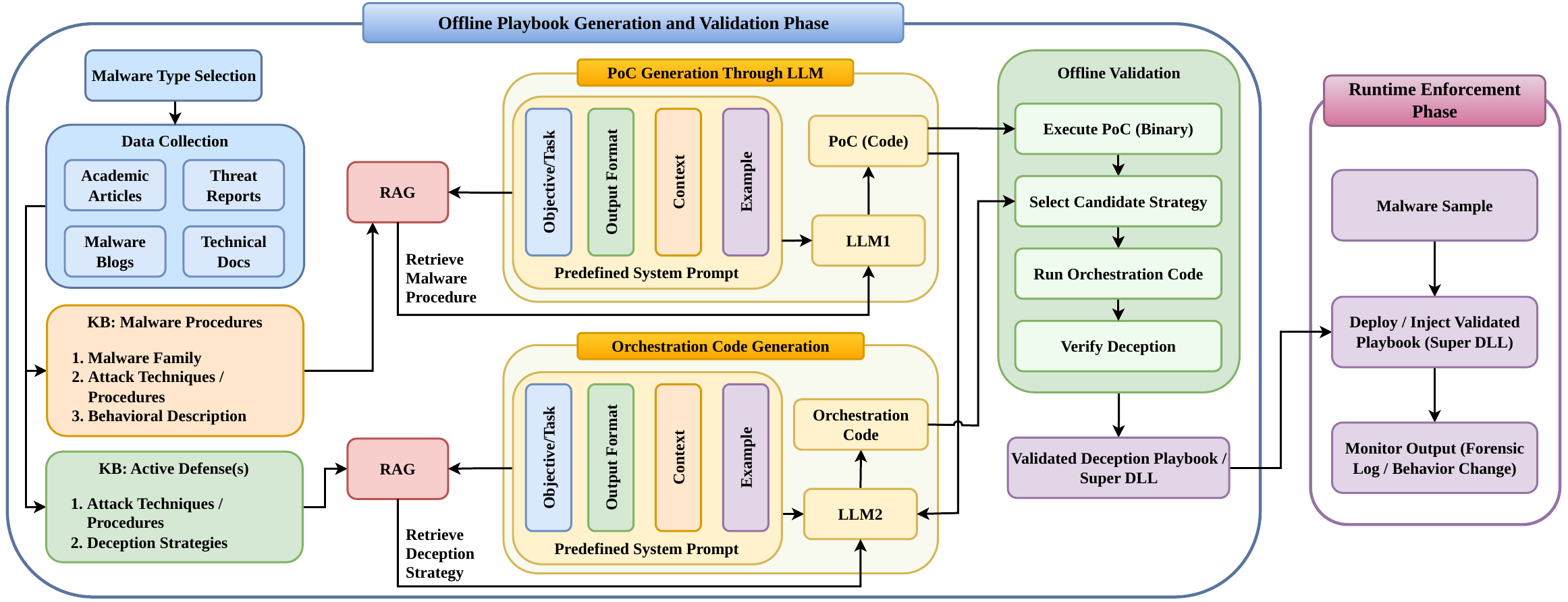}
    \vspace{-0.2cm}
    \caption{ORCAGen architecture for offline playbook generation and runtime deception enforcement. LLM1 and LLM2 denote separate generation roles for PoC malware and deception orchestration code.}
    \vspace{-0.4cm}
    \label{fig:system_architecture}
\end{figure*}

\subsection{Deception-Oriented System Orchestration}
Cyber deception aims to mislead malicious activity while preserving defender visibility. In endpoint settings, this can include presenting fake files, redirecting credential access, altering API responses, or returning controlled outputs that cause malware to operate on decoy resources instead of real assets. Orchestration refers to the process of selecting and deploying these deception actions based on observed behavior. The key challenge is that deception logic must be both context-specific and safe to enforce. Static playbooks are easy to deploy but difficult to adapt to new malware behaviors. Fully adaptive runtime logic may respond to changing behavior, but can introduce latency, unpredictability, or operational risk if the decision process is not validated. ORCAGen follows a middle path: it uses GenAI and RAG to construct malware-specific deception logic offline, validates that logic in a controlled setting, and then deploys only the verified playbook for deterministic runtime enforcement.

\section{Threat Model}\label{sec:threat_model}
ORCAGen targets a post-compromise endpoint setting in which malware has reached execution and begins interacting with user-level resources before traditional defenses fully contain it. {ORCAGen is not a malware detector, remover, or access-control mechanism. Instead, it assumes that candidate suspicious processes are supplied by an external signal, such as an endpoint detector, sandbox classification, analyst rule, or deployment policy. ORCAGen then applies pre-validated deception logic only to suspicious or policy-selected processes, not to all system processes. In this work, we focus on user-space malware, including keyloggers, information stealers, and ransomware, because these families expose observable interactions with keyboard input, clipboard data, screen capture APIs, credential files, file-system traversal, registry entries, and file read/write operations. ORCAGen assumes that such interactions pass through userland APIs or user-level artifacts that can be monitored and manipulated through deception.

We assume the defender has sufficient endpoint control to deploy ORCAGen's runtime enforcement module, inject the validated Super DLL into suspicious processes, monitor selected API and file-system events, and maintain whitelists of trusted applications. ORCAGen separates playbook construction from enforcement: RAG-guided LLMs generate and validate deception logic offline, while the Super DLL enforces only pre-validated responses during malware execution. Malware that operates entirely outside monitored user-space behaviors, runs with kernel-level privileges, or does not interact with protected resources is outside the scope of this evaluation. Anti-hooking and defensive-hook tampering behaviors~\cite{sajid2025secure} can be incorporated when corresponding procedures and countermeasures are added to the knowledge base, but they are not the focus of this study. ORCAGen protects selected user-level assets and preserves forensic visibility; it does not replace endpoint detection, access control, backup, or malware removal.

%% file: IEEE-TPS/file-methodology.tex
\section{ORCAGen: System Overview}\label{sec:methodology}
\subsection{System Architecture}

A deception playbook is a validated set of behavior triggers, monitored APIs or artifacts, intervention points, and deception responses for a target malware behavior. In ORCAGen, each playbook maps observed malware actions to pre-tested interventions, such as redirecting credential access, suppressing sensitive data collection, manipulating API outputs, or returning fake success or failure. The key insight is to generate and validate malware-specific deception logic offline, then enforce only pre-validated actions at runtime. This avoids LLM inference during malware execution, reducing latency, nondeterministic behavior, and unsafe runtime decisions. The tradeoff is that runtime enforcement is limited to behaviors and interventions represented in the KB and validated before deployment. 

ORCAGen therefore builds deception playbooks in one phase and enforces them in another. Figure~\ref{fig:system_architecture} summarizes this workflow. During offline playbook construction, ORCAGen takes a target malware family or behavior, optionally supplemented with analyst-provided logs or malware analysis notes, retrieves relevant malware and defense knowledge through RAG, generates PoC malware and deception orchestration code, and validates their interaction in a controlled environment. Only code that passes validation is added to the final playbook. At runtime, ORCAGen does not query an LLM; it injects the validated Super DLL into suspicious or policy-selected processes and enforces pre-tested deception responses.

\subsection{Knowledge Base Creation for RAG}\label{kb}

ORCAGen uses two structured knowledge bases (KBs) to ground generation in concrete malware behavior and corresponding deception actions. The first KB, \emph{Malware Procedures}, captures how a malware family performs a target behavior, including the attack technique, behavioral description, relevant APIs or artifacts, and observable execution pattern. The second KB, \emph{Active Defenses}, maps those behaviors to deception strategies that can redirect, suppress, or mislead the malware. In this work, we focus on keyloggers, information stealers, and ransomware because their behaviors expose clear user-level interactions, such as keystroke capture, credential access, screen capture, file-system traversal, and file encryption. These families provide a concrete starting point for evaluating ORCAGen, while the same KB structure can be extended to additional malware families by adding new procedure and defense entries.

\begin{figure}[!ht]
    \centering
    \vspace{-0.1cm}
    \includegraphics[width=.85\columnwidth]{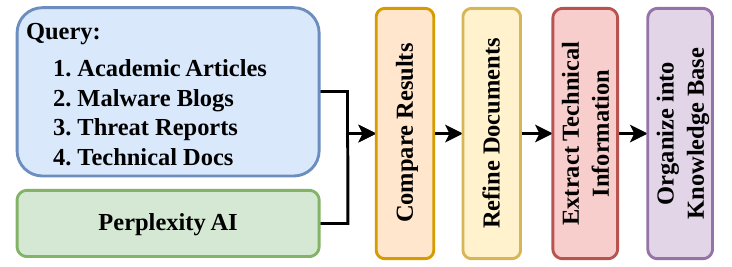}
    \vspace{-0.3cm}
    \caption{Knowledge Base Creation Process.}
    \label{fig:kb_creation}
    \vspace{-0.2cm}
\end{figure}

To populate the KBs, we used the collection workflow shown in Figure~\ref{fig:kb_creation}. The pipeline combines lexical search, semantic search, and manual validation. Seed queries were manually curated from the target malware families, relevant MITRE ATT\&CK techniques, and known APIs or artifacts associated with those behaviors. For example, \texttt{GetAsyncKeyState} is a Windows API commonly abused by keyloggers to poll keyboard state, so it was used as a seed indicator for keylogging-related searches. Exact-phrase queries, such as ``GetAsyncKeyState keylogger," were used to retrieve sources with precise API-level discussions, while flexible multi-term queries, such as \texttt{GetAsyncKeyState + keylogging + evasion}, were used to capture sources where relevant concepts appear in different orders or contexts. We seeded the search process with MITRE ATT\&CK techniques, including \texttt{T1056: Input Capture}, \texttt{T1003: Credential Dumping}, and \texttt{T1486: Data Encrypted for Impact}, and expanded the query set with deception-oriented terms such as ``API hooking mitigation," ``clipboard decoy," ``fake credential injection," and ``bait files." 
Candidate documents were collected from academic papers, threat intelligence reports, technical documentation, and malware blogs. We also used semantic search to capture contextually relevant sources that did not share exact keywords with the seed queries. Sources that appeared across both lexical and semantic retrieval were prioritized for review because they provided stronger topical evidence. The repository includes retained source metadata, query seeds, and mappings from source documents to KB entries. A detailed distribution of collected and retained documents is provided in Appendix~\ref{app:KB_distribution}.

After collection, retrieved documents were manually filtered and converted into structured entries.
Each entry contains four core fields: i) malware family, which identifies the broad threat class; ii) attack technique, which specifies the behavior to model; iii) behavioral description, which records how the behavior is implemented through APIs, artifacts, or system interactions; and iv) deception strategy, which defines the corresponding active-defense response. We encode each behavior at two complementary levels. At the \emph{procedure level}, entries capture how malware performs an action in practice, including concrete APIs, artifacts, system interactions, and execution patterns, such as \texttt{GetAsyncKeyState}, \texttt{SetWindowsHookEx}, browser credential stores, registry entries, or file-system APIs. 
Artifacts refer to observable system objects or data sources touched by malware, such as credential files, browser profiles, registry keys, clipboard contents, screenshots, file paths, or bait files. At the \emph{response level}, entries capture the corresponding defender-side intervention, including the deception objective, hook or monitor, trigger condition, safety constraint, and expected response, such as injecting false keystrokes, redirecting credential access to honey artifacts, returning decoy file names, or simulating success or failure for destructive file operations. This layered structure is important for ORCAGen because PoC generation requires low-level behavioral details, while orchestration generation requires defense strategies that can generalize across related malware behaviors. The final KBs are stored in structured JSON format and indexed for retrieval during RAG. A representative KB example is provided in Appendix~\ref{app:kb_example}, while the complete schema, query seeds, source metadata, and additional KB entries are provided in~\cite{orcagen}.

\subsection{RAG Implementation}\label{rag_impl}

The RAG component controls what threat and defense knowledge is exposed to the LLM during offline playbook construction. Rather than providing the full knowledge base to the model, ORCAGen retrieves only the entries relevant to the current generation task and injects them into the \emph{Context} field of the prompt template described in Section~\ref{prompt_egnr}. Malware procedure entries provide the target family, attack technique, behavioral description, relevant APIs or artifacts, affected resources, and expected execution pattern. Active defense entries provide the deception objective, intervention point, required hook or monitor, expected response, and safety constraints. In the current prototype, the KB contains \textbf{8} malware procedure entries and \textbf{19} active defense entries covering \textbf{3} malware families, \textbf{8} attack techniques, and \textbf{28} API-level or file-system indicators. For each generation task, ORCAGen forms a retrieval query from the malware family, target behavior, attack technique, and generation objective. As the KB is stored as structured JSON records, ORCAGen retrieves at the entry level rather than using arbitrary fixed-length text chunks. Each malware procedure or active-defense record is indexed as one retrieval unit, and the top-\textbf{k} matching entries are inserted into the prompt context. For PoC generation, the retrieved context describes the malware behavior to emulate, including relevant APIs, artifacts, and execution patterns. For orchestration generation, the retrieved context describes the corresponding active defense strategy and is combined with the generated PoC code. This separation grounds the PoC generator on malware behavior(s) and the orchestration generator on defensive intervention logic.

ORCAGen supports two RAG modes. For cloud-hosted LLMs, including ChatGPT, Gemini, and Claude, we used the platforms' built-in knowledge or file-grounding mechanisms to make the KB available during generation. For local models, such as Qwen through Ollama and Open WebUI, we used an external lightweight RAG pipeline that indexes the structured KB and retrieves the top-\textbf{k} entries for prompt injection. Across both modes, the workflow remains the same: retrieve relevant threat and defense entries, insert the retrieved content into the structured prompt, and generate either PoC malware or deception orchestration code. This hybrid setup enables ORCAGen to run across both cloud-hosted and local LLM backends, but it also affects reproducibility. Cloud platforms do not expose full control over their internal retrieval mechanisms, so our cross-model results should be interpreted as a comparison of practical LLM-backend configurations rather than a controlled comparison of model weights alone. To address this limitation, our planned deployment uses a unified external RAG layer for all models. In that design, ORCAGen performs retrieval externally, logs the query and retrieved KB entries, constructs the final prompt, and sends it to online models through API-based invocation. The retrieval and prompt-construction pipeline therefore remains consistent across local and cloud-hosted models; only the communication interface changes.

\subsection{Prompt Engineering}\label{prompt_egnr}
\begin{figure*}[!ht]
    \centering
    \includegraphics[width=.86\textwidth] {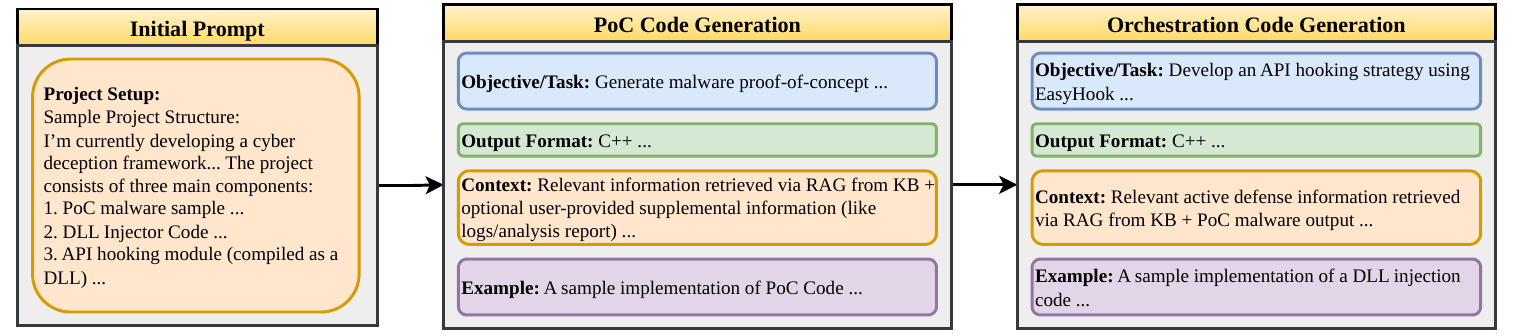}
    \vspace{-0.3cm}
    \caption{Formal Prompt Structure.}
    \vspace{-0.4cm}
    \label{fig:prompt_structure}
\end{figure*}

Prompt engineering in ORCAGen is used to make LLM outputs executable, testable, and aligned with the deception workflow. A generic prompt is often insufficient for this setting because the model must produce code that fits a specific project structure, uses compatible APIs, and interacts correctly with the generated PoC or runtime hook. ORCAGen therefore uses structured prompt templates that specify the task, output constraints, retrieved context, and examples needed for malware-specific generation. Figure~\ref{fig:prompt_structure} shows the prompt structure used for PoC and orchestration code generation.

\textbf{Initial Prompt:} ORCAGen begins with an initial project-level prompt that defines the expected code organization. The prompt describes three components: a controlled PoC malware sample, a DLL-based API hooking module, and an injector for deploying the DLL into a target process. It also includes partial code snippets and implementation constraints so that the model understands the expected execution flow and interface between components. This scaffold is important because, without it, LLMs often generate incomplete projects, mismatched function signatures, incompatible character encoding, or code fragments that cannot be compiled or tested. Providing the scaffold constrains the output space and makes the generated artifacts easier to validate.

\textbf{Prompt Template Components:} Each generation prompt contains four components: objective/task, output format, context, and example. The \emph{objective/task} defines the generation goal, such as producing a PoC for a target malware behavior or producing deception orchestration code for a given PoC. The \emph{output format} specifies implementation constraints, including the programming language, required libraries, function definitions, and expected file structure. The \emph{context} field contains retrieved knowledge from the RAG pipeline, as well as any optional user-provided artifacts such as malware analysis notes, execution logs, or forensic observations. ORCAGen does not manually paste the full KB into this field; instead, relevant entries are retrieved and injected for the current task. The \emph{example} field provides minimal code snippets or working patterns that guide the model toward syntactically valid and compatible outputs. Together, these components help reduce hallucinated APIs, incomplete logic, and outputs that do not match the deployment environment.

\textbf{Two-Stage Prompt Use:} ORCAGen applies the template in two generation stages. In the first stage, the prompt is grounded with malware procedure entries and asks the model to generate a controlled PoC that reproduces a specific behavior, such as keylogging, credential access, file scanning, or ransomware-style file modification. In the second stage, the generated PoC is combined with active defense entries retrieved from the KB, and the model is asked to generate deception orchestration code. This separation keeps the malware-emulation task and the defense-generation task distinct while allowing the orchestration prompt to reason over the behavior exposed by the PoC.

\textbf{Prompt Refinement:} If the initial output does not compile or fails to produce the intended behavior, ORCAGen uses iterative refinement prompts, with no human edits to source code are allowed. These prompts provide targeted feedback, such as compiler errors, missing functionality, incorrect API usage, or runtime observations, while preserving previously valid components. For example, if the generated orchestration code lacks DLL injection logic or uses an incompatible function signature, the refinement prompt asks the model to correct that specific part rather than regenerate the entire project. This process continues until runnable PoCs and valid orchestration code are produced or the refinement limit is reached. The number of refinements required is recorded as an evaluation metric because it reflects how much post-processing each LLM requires for this domain-specific task.

\subsection{Code Generation}
The code generation module produces the two artifacts needed to construct a deception playbook: controlled PoC malware and corresponding deception orchestration code. Both artifacts are generated during the offline phase using the structured prompt template and the RAG-retrieved context described earlier. The purpose of generating PoC malware is not to create deployable malware, but to obtain a controlled and reproducible instance of a target behavior that can be used to validate the deception logic before runtime deployment.

\textbf{PoC malware generation:} Given a target malware family, such as a keylogger, information stealer, or ransomware, ORCAGen retrieves malware procedure entries from the KB and injects them into the PoC generation prompt. These entries describe the behavior to emulate, including relevant APIs, artifacts, execution patterns, and affected resources. The LLM then generates PoC code that reproduces the selected behavior in a controlled form, such as keyboard capture, credential-file access, file-system scanning, or ransomware-style file modification. The generated PoC is compiled and executed in a controlled environment to confirm that it exposes the intended behavior. PoCs that do not compile or fail to reproduce the target behavior are refined or discarded.

\textbf{Orchestration code generation:} After the PoC is generated, ORCAGen retrieves the corresponding active defense entries from the KB. These entries describe the deception objective, intervention point, required hook or monitor, expected response, and safety constraints. The orchestration generation prompt receives both the generated PoC code and the retrieved defense context. The LLM then produces deception-oriented orchestration code, such as API hooks, file-system redirection logic, fake-success or fake-failure responses, decoy credential redirection, or screen and input manipulation. Depending on the retrieved defense strategy, ORCAGen may generate multiple orchestration variants for the same malware behavior. These variants are not directly deployed; they are first compiled, reviewed, and validated against the controlled PoC behavior. Only validated orchestration logic is eligible for inclusion in the final playbook.

\begin{figure*}
    \centering
    {\includegraphics[width=0.42\textwidth]{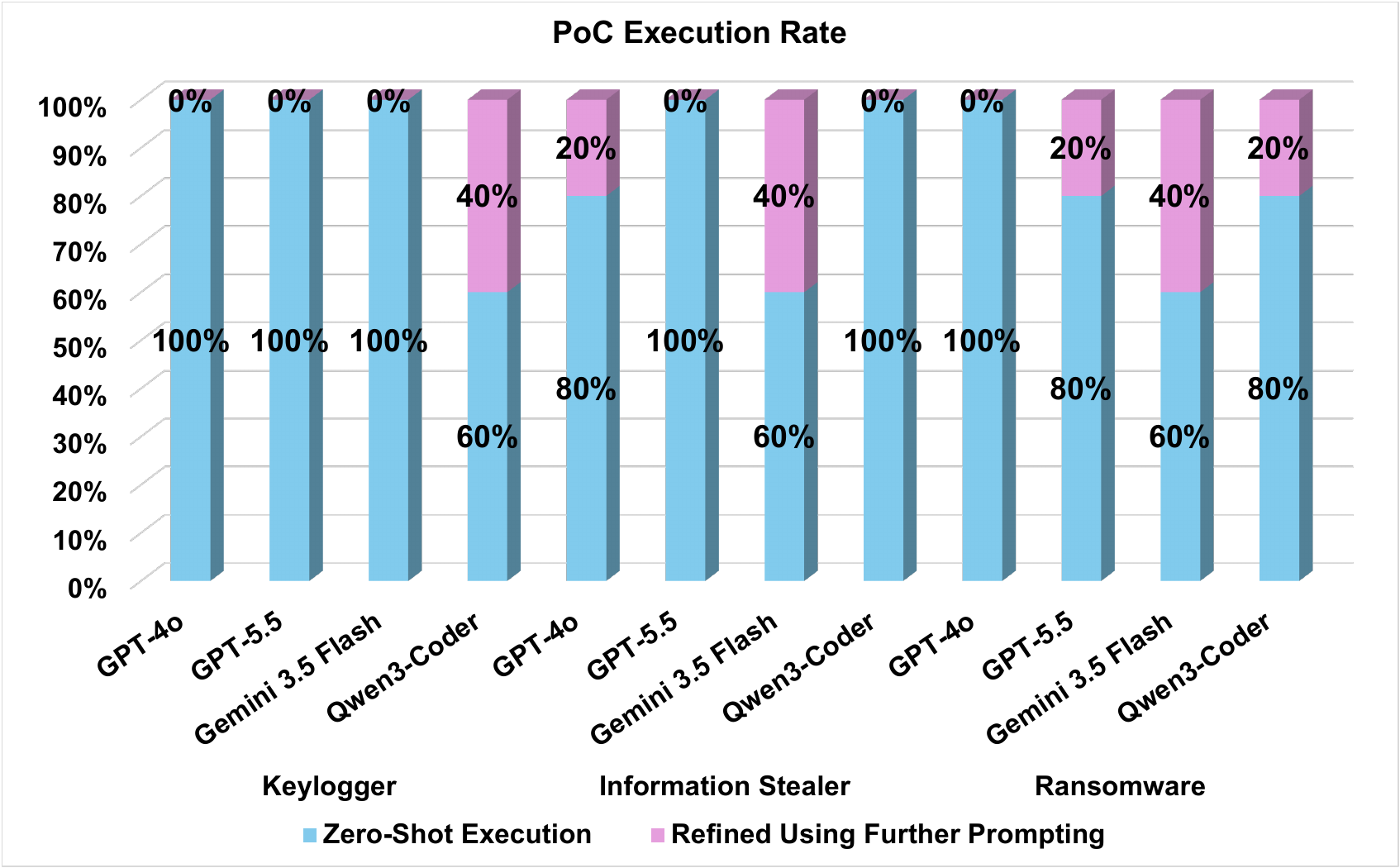}} 
    {\includegraphics[width=0.42\textwidth]{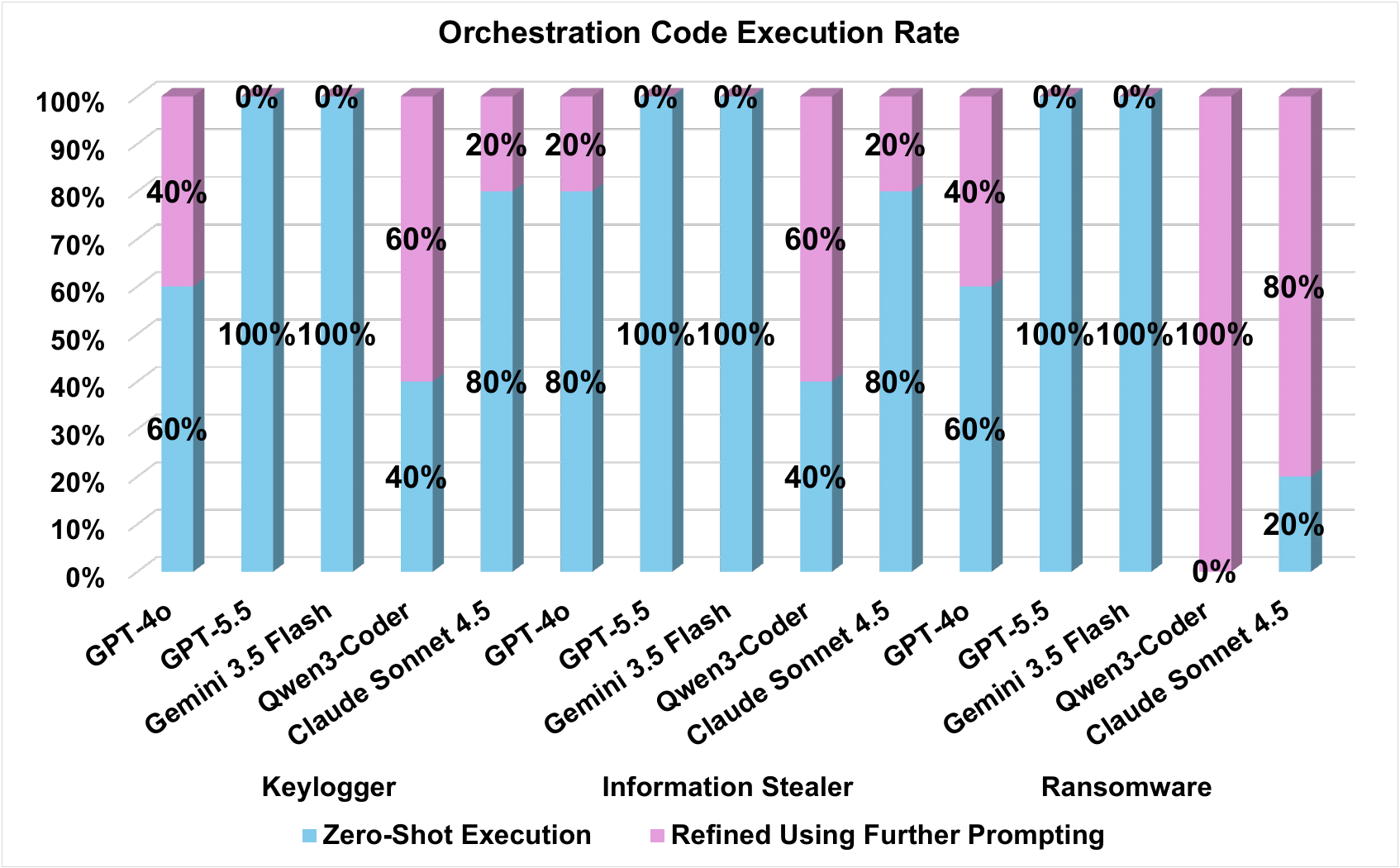}} 
    \vspace{-0.25cm}
    \caption{Functionality Validation: (a) PoC Execution Rate, (b) Orchestration Code Execution Rate.}
    \vspace{-0.45cm}
    \label{fig:functionality_validation}
\end{figure*}

\subsection{Offline Validation and Runtime Enforcement}

ORCAGen validates generated deception logic before adding it to the runtime playbook. After a PoC and its corresponding orchestration code are generated, the PoC is first executed in a controlled environment to verify that it reproduces the intended behavior, such as credential theft, keystroke logging, file scanning, screen capture, or file encryption. The orchestration code is then compiled and executed against the PoC to determine whether it can redirect, delay, suppress, or mislead the malware behavior without disrupting benign activity in the same test environment. ORCAGen records execution traces, forensic logs, behavioral changes, false positives, and false negatives during this validation stage. These results determine whether the generated deception strategy is reliable enough to be added to the playbook.

Only deception strategies that pass validation are compiled into the reusable playbook, implemented as a Super DLL. ORCAGen does not replace malware detection; it assumes a candidate process has been flagged by an endpoint detector, sandbox, analyst rule, or deployment policy. During runtime enforcement, the Super DLL is injected into that suspicious or policy-selected process. Deception responses are gated by process trust status and behavior-specific triggers, allowing benign or whitelisted processes to proceed normally. 
The Super DLL monitors malware-relevant API calls, file operations, credential-access attempts, input-capture behavior, and other indicators covered by the playbook. When a monitored behavior matches a validated strategy, the Super DLL triggers the corresponding deception response and records the resulting logs and behavioral changes. This phase evaluates the runtime behavior of the playbook, not live LLM inference, since all generation, refinement, and validation have already been completed offline.

%% file: IEEE-TPS/file-eval.tex
\section{Evaluation}\label{sec:evaluation}

This section evaluates ORCAGen across its two-phase workflow. First, we evaluate the offline generation and validation phase by measuring whether different LLMs can generate functional proof-of-concept (PoC) malware and corresponding deception orchestration code. Second, we evaluate the runtime enforcement phase by compiling validated orchestration logic into a deception playbook and testing it against synthesized and real-world malware samples. We evaluated five LLM backends: \textbf{GPT-4o, GPT-5.5, Gemini 3.5 Flash, Qwen3-Coder (30B), and Claude Sonnet 4.5}. Code generation was conducted on a Windows 11 Enterprise host equipped with a 13th Gen Intel(R) Core i7-13700 CPU, 32 GB RAM, and an NVIDIA GeForce RTX 3070 GPU. Generated code was compiled and tested in VMware Workstation Pro 17 using an isolated Windows 7 virtual machine configured with 8 GB RAM. We selected Windows 7 because it provides a lightweight and controllable environment for evaluating user-level Windows API interactions while reducing virtualization overhead. Although Windows 10 and Windows 11 include newer platform protections, the API-level behaviors studied in this work, including input capture, file-system traversal, credential-file access, and file I/O operations, remain suitable for controlled validation in this testbed. All PoC malware and orchestration logic were executed only inside the isolated virtual environment.

\subsection{Functionality Validation}
This subsection evaluates whether ORCAGen can generate functional PoC malware and corresponding deception orchestration code before adding the generated logic to the runtime playbook. The evaluation has two objectives: (i) determine whether the generated PoC reproduces the targeted malicious behavior, and (ii) determine whether the generated orchestration code executes correctly and disrupts, redirects, or conceals that behavior in the controlled testbed.

\textbf{Setup:} We evaluated three malware categories: keyloggers, information stealers, and ransomware. For each category, we designed five simulation scenarios, resulting in fifteen total test cases. In each scenario, ORCAGen was tasked with generating a PoC malware sample and a corresponding orchestration module. The PoC was expected to reproduce a specific behavior, such as keystroke capture, clipboard capture, credential-file access, file scanning, or ransomware-style file modification. The orchestration module was expected to intervene through the corresponding deception strategy, such as redirecting access, returning decoy artifacts, suppressing sensitive output, or simulating operation success or failure. We monitored execution traces, logs, and observable behavior to confirm whether each PoC and orchestration module performed the intended function. The full list of simulation scenarios is available in our repository~\cite{orcagen}.

\textbf{Metrics:} We report two primary metrics. \textbf{PoC execution rate} measures the percentage of generated PoCs that compiled, executed, and reproduced the expected malicious behavior. \textbf{Orchestration code execution rate} measures the percentage of generated orchestration modules that compiled, executed, and successfully intervened against the corresponding PoC. We further classify each output into three categories: \textbf{zero-shot execution}, where the generated code worked on the first attempt; \textbf{minor correction}, where the code required straightforward fixes before successful execution; and \textbf{major reliability issue}, where the output contained structural or semantic flaws that prevented reliable validation. Minor corrections included missing headers, ANSI/Unicode mismatches, C++ compatibility adjustments, and simple type-casting errors. Major reliability issues included hallucinated APIs, fabricated system calls, logically impossible behavior, or incomplete code structure.


\textbf{Results and Discussion:}
Figure~\ref{fig:functionality_validation} summarizes PoC and orchestration execution rates across LLMs. For PoC generation, GPT-5.5 and GPT-4o achieved the strongest overall reliability, with 93.33\% zero-shot execution and only one minor correction. Gemini 3.5 Flash achieved 100\% zero-shot success for keyloggers but dropped to 60\% for information stealers and ransomware. Qwen3-Coder showed mixed results, with 60\%, 100\%, and 80\% zero-shot success for keyloggers, information stealers, and ransomware, respectively. Claude Sonnet 4.5 did not generate PoC malware because of content safety restrictions and was evaluated only for orchestration using GPT-5.5-generated PoCs. For orchestration code generation, GPT-5.5 and Gemini 3.5 Flash achieved 100\% zero-shot execution across all malware categories. GPT-4o achieved 66.67\% zero-shot orchestration execution overall, while Qwen3-Coder required the most refinements, reaching 40\% zero-shot success for keyloggers and information stealers and requiring additional prompting for all ransomware cases. Claude Sonnet 4.5 achieved 80\% zero-shot orchestration execution for keyloggers and information stealers, but only 20\% for ransomware.
Table~\ref{tab:error_categories} summarizes the observed errors. GPT-5.5 produced the fewest issues, with one compatibility-related correction and no hallucinated APIs, while Qwen3-Coder produced the most errors, including four hallucinated API or data references. GPT-4o, Gemini 3.5 Flash, and Claude Sonnet 4.5 mostly produced minor implementation errors, such as missing headers, encoding mismatches, type-casting issues, or C++ compatibility problems. Overall, these results show that LLMs can generate functional PoC and deception orchestration code, but reliability differs substantially across models and tasks, reinforcing the need for ORCAGen's offline validation before generated logic is added to the runtime playbook.

\begin{table}[h]

\centering
\renewcommand{\arraystretch}{1.2} 
\setlength{\tabcolsep}{4pt} 
\scriptsize 
\begin{tabular}{|p{1.9cm}|p{0.9cm}|p{1.1cm}|p{1.2cm}|p{0.9cm}|p{1.25cm}|}
\hline
\textbf{GenAI Model} & \textbf{Missing Headers} & \textbf{Encoding Mismatch} & \textbf{C++ Compatibility} & \textbf{Type Casting} & \textbf{Hallucinated APIs/Data} \\ \hline
GPT-4o & 1 & 5 & 0 & 0 & 0 \\ \hline
GPT-5.5 & 0 & 0 & 1 & 0 & 0 \\ \hline
Gemini 3.5 Flash & 1 & 0 & 2 & 1 & 0 \\ \hline
Qwen3-Coder & 8 & 3 & 1 & 1 & 4 \\ \hline
Claude Sonnet 4.5 & 3 & 0 & 1 & 1 & 0 \\ \hline
\end{tabular}
\vspace{1mm}
\caption{Validation Errors Observed Across LLM Backends}
\vspace{-8mm}
\label{tab:error_categories}
\end{table}

\subsection{Orchestration Accuracy}

\subsubsection{Against Synthesized Malware Samples}
This experiment evaluates whether ORCAGen-generated orchestration code can successfully deceive or neutralize the PoC malware generated during offline validation. The evaluation uses the same three malware families as the functionality validation experiment: keyloggers, information stealers, and ransomware. For each family, we use five simulation scenarios, resulting in fifteen test cases. Each test case includes a generated PoC and its corresponding orchestration code, both executed in a controlled virtual environment. The goal is to determine whether the orchestration logic can alter, suppress, redirect, or conceal the targeted malicious behavior without disrupting legitimate system functionality.

The primary metric is the \textbf{deception success rate}, defined as the percentage of executions in which the orchestration code successfully changed the outcome of the malicious behavior. We also track the number of \textbf{prompt refinements} required for each model to reach full success. A refinement corresponds to a targeted follow-up prompt that corrects missing logic, compilation issues, or runtime behavior observed during validation. For this experiment, we allowed a maximum of three prompt refinements per scenario. To assess safety within the controlled validation setting, we measure two failure conditions: \textbf{false positives}, where a validated orchestration response affects a process interaction that should be allowed under the playbook policy, and \textbf{false negatives}, where the orchestration fails to intervene and allows the malware behavior to complete. 
As the ORCAGen-generated Super DLL is injected only into suspicious or policy-selected processes, this metric evaluates whether the playbook applies its deception response too broadly within the monitored process, not whether ORCAGen disrupts arbitrary benign applications on the host.

\begin{figure}[!ht]
    \centering
    \includegraphics[width=.95\columnwidth]{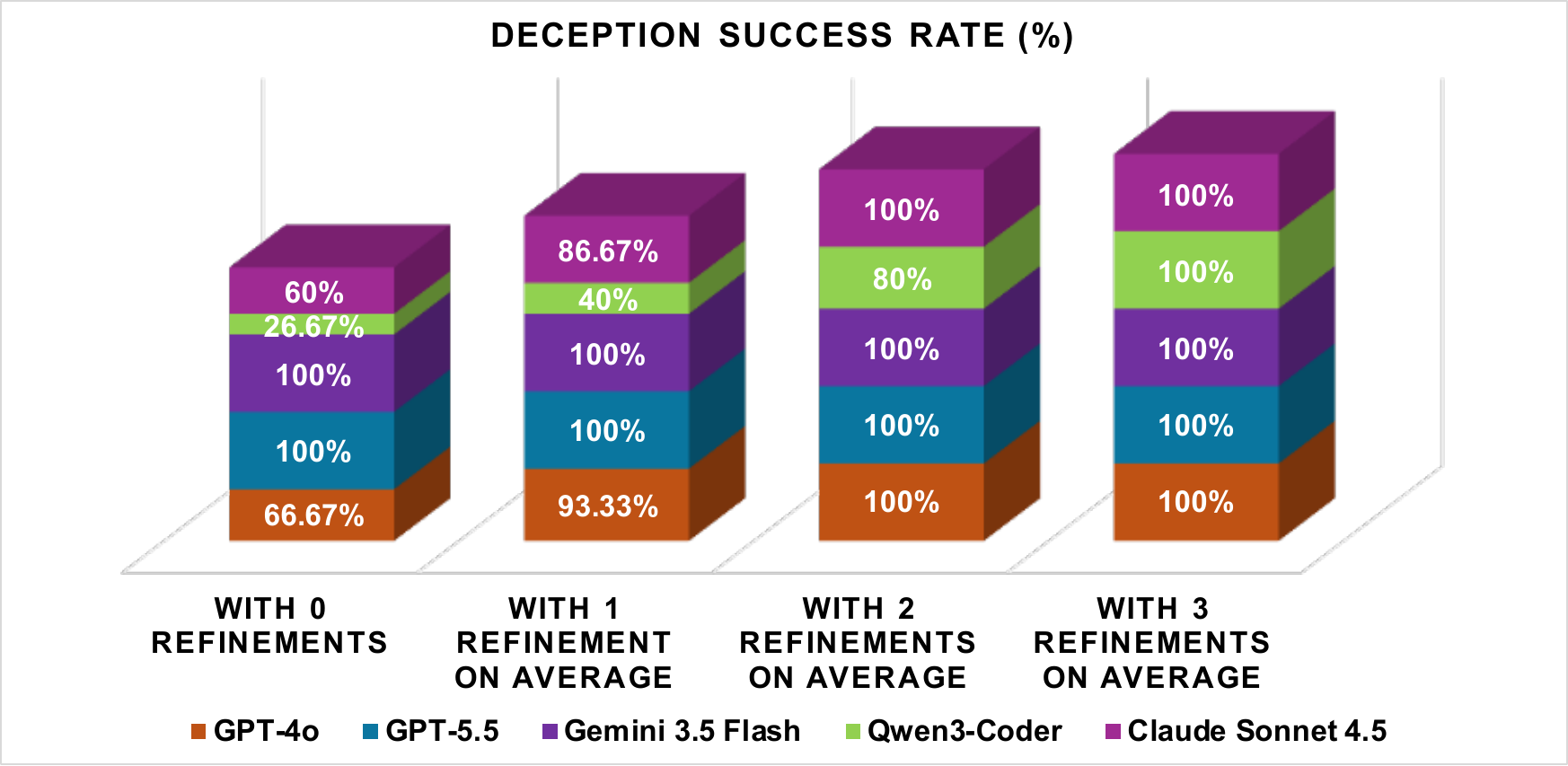}
    \vspace{-0.25cm}
    \caption{Orchestration Accuracy of GenAI Models.}
    \vspace{-0.5cm}
    \label{fig:orchestration_accuracy}
\end{figure}

\textbf{Results and Discussion:} 
Figure~\ref{fig:orchestration_accuracy} shows deception success rates as prompt refinements are introduced. All models eventually reached 100\% success across the synthesized malware scenarios, showing that ORCAGen can produce correct orchestration for selected malware families when refinement is allowed. However, zero-shot performance varied substantially. GPT-5.5 and Gemini 3.5 Flash achieved 100\% zero-shot deception success, while GPT-4o, Claude Sonnet 4.5, and Qwen3-Coder achieved 66.67\% (10/15), 60\% (9/15), and 26.67\% (4/15), respectively. With refinement, GPT-4o and Claude Sonnet 4.5 reached 100\% success after an average of two refinements, while Qwen3-Coder required three. These results indicate that generated orchestration logic is generally correctable, but post-processing effort depends strongly on the LLM backend. Across all synthesized experiments, we observed no false positives or false negatives after validation. Within the monitored validation setting, the validated orchestration logic did not affect interactions allowed under the playbook policy and consistently altered or neutralized the targeted PoC behavior. These results support ORCAGen's offline validation design: even when refinement is required, the final playbook includes only logic tested against the target behavior and checked for unintended disruption.

\textbf{Ethics and Safety:} ORCAGen generates PoC malware only to validate deception logic in a controlled environment. All PoCs were behavior-limited and executed inside isolated virtual machines. Ransomware PoCs operated only on dummy files, credential-theft behavior was redirected to synthetic artifacts, and network exfiltration was disabled or redirected to local sinks. We do not release weaponizable malware code. Released artifacts include sanitized prompts, KB schemas, benign test harnesses, and evaluation metadata.


\subsubsection{Against Real-World Malware Samples}
We next evaluate whether ORCAGen’s validated playbooks transfer beyond synthesized PoC malware. This evaluation is independent of the LLM-generated PoCs and tests the playbooks against 150 real-world malware samples with 50 samples each from keyloggers, information stealers, and ransomware. The samples were collected from ANY.RUN and VirusTotal between March and May 2026. The dataset includes multiple subfamilies within each category, and the complete subfamily list and per-sample evaluation metadata are provided in our repository~\cite{orcagen}. Each sample was executed in a controlled and instrumented environment, and the corresponding pre-generated deception playbook was deployed to determine whether it could alter, suppress, or safely redirect the malware's intended behavior.

\textbf{For information stealers}, ORCAGen protected credential-access targets commonly abused by malware, including browser credential stores, FTP credentials, and application-specific credential files used by email clients. Following target categories identified in prior work~\cite{sajid2020dodgetron}, ORCAGen monitored file operations on these sensitive targets and checked whether the accessing process was absent from a predefined whitelist of legitimate applications. When both conditions were satisfied, the process was treated as suspicious and redirected to honey credential files instead of real credential artifacts. 
If an upstream detector, sandbox, analyst rule, or deployment policy incorrectly selects a benign process for injection, ORCAGen enforces the active playbook policy on that process. The resulting interaction logs allow an analyst to review the event and update the whitelist or process-selection policy when the behavior is legitimate.

\textbf{For ransomware}, ORCAGen used two complementary deception strategies inspired by prior bait-based ransomware defenses~\cite{sajid2025randecepter}. First, bait files were exposed through soft links at the top and bottom of monitored directories to reduce storage overhead while preserving coverage. If a bait file was encrypted, ORCAGen identified the responsible process and marked it as ransomware when the process was not part of the trusted whitelist. Second, ORCAGen monitored destructive file operations, including deletion and overwrite attempts, and returned deceptive success or failure responses while preserving the original files. These interventions were applied only to selected high-value user locations and excluded temporary folders or other low-value directories where benign applications frequently perform high-volume file modifications.

\textbf{For keyloggers}, ORCAGen focused on input, clipboard, and screen-capture behaviors. It monitored suspicious attempts to capture keyboard input, access clipboard content, or collect screen and cursor information. For keyboard and clipboard logging, ORCAGen suppressed or replaced captured content with decoy input or honey clipboard data. For screen logging, it redirected capture attempts toward decoy visual information, preventing the malware from observing the actual user screen state. These interventions were applied only to suspicious processes to avoid disrupting benign applications that legitimately use input, clipboard, or screen-capture functionality. Table~\ref{tab:real_malware_accuracy} summarizes the real-world malware results. We classify each execution into three categories: \textbf{complete neutralization}, where the malware's intended behavior was fully disrupted or redirected; \textbf{partial disruption}, where only part of the malicious behavior was affected or observable; and \textbf{deception failure}, where the playbook produced no observable defensive effect.

\begin{table}[h]
\centering
\renewcommand{\arraystretch}{1.3}
\setlength{\tabcolsep}{4pt}
\scriptsize
\begin{tabular}{|p{2.0cm}|p{1.8cm}|p{1.5cm}|p{1.5cm}|}
\hline
\textbf{Malware Family} & \textbf{Complete Neutralization} & \textbf{Partial Disruption} & \textbf{Deception Failure} \\ \hline
Keylogger & 92\% & 0\% & 8\% \\ \hline
Information Stealer & 100\% & 0\% & 0\% \\ \hline
Ransomware & 96\% & 2\% & 2\% \\ \hline
\end{tabular}
\vspace{2mm}
\caption{ORCAGen Orchestration Accuracy against Real Malware}
\vspace{-4.5mm}
\label{tab:real_malware_accuracy}
\end{table}

\textbf{Results and Discussion:}
ORCAGen achieved strong effectiveness across all three real-world malware families. It completely neutralized 92\% of keyloggers, 100\% of information stealers, and 96\% of ransomware samples. The four failed keylogger samples used direct keyboard-state polling with internal key mapping; ORCAGen could suppress or simulate key-state responses, but could not consistently replace internally reconstructed key values with decoy keystrokes. For ransomware, the partial case depended on a command-and-control server to download an additional component; because the infrastructure was unavailable during testing, only part of the behavior was observed, so we conservatively classified it as partial disruption. Overall, these results suggest that ORCAGen's playbooks can transfer from synthesized PoC validation to real-world malware when malicious behavior interacts with monitored input, credential, file-system, or API-level targets.

\subsection{Ablation and Baseline Comparison}
This subsection evaluates whether ORCAGen's performance comes from the complete framework or simply from repeated prompting. We conduct two comparisons. First, we compare three prompting strategies using the same 15 scenarios from the functionality validation experiment: \textbf{Direct Prompting}, \textbf{RAG-Enhanced Prompting}, and \textbf{ORCAGen}. Direct Prompting asks the LLM to generate deception orchestration code without external context. RAG-Enhanced Prompting adds the retrieved deception strategy from the KB but does not use ORCAGen's full prompt structure. ORCAGen combines retrieved defense knowledge with structured prompt components, including the objective, context, output format, examples, and validation-oriented project scaffold. All three strategies were evaluated with GPT-5.5 using the same refinement procedure, with a maximum of five prompt refinements per scenario.

\begin{figure}[!ht]
    \centering
    \includegraphics[width=0.95\columnwidth]{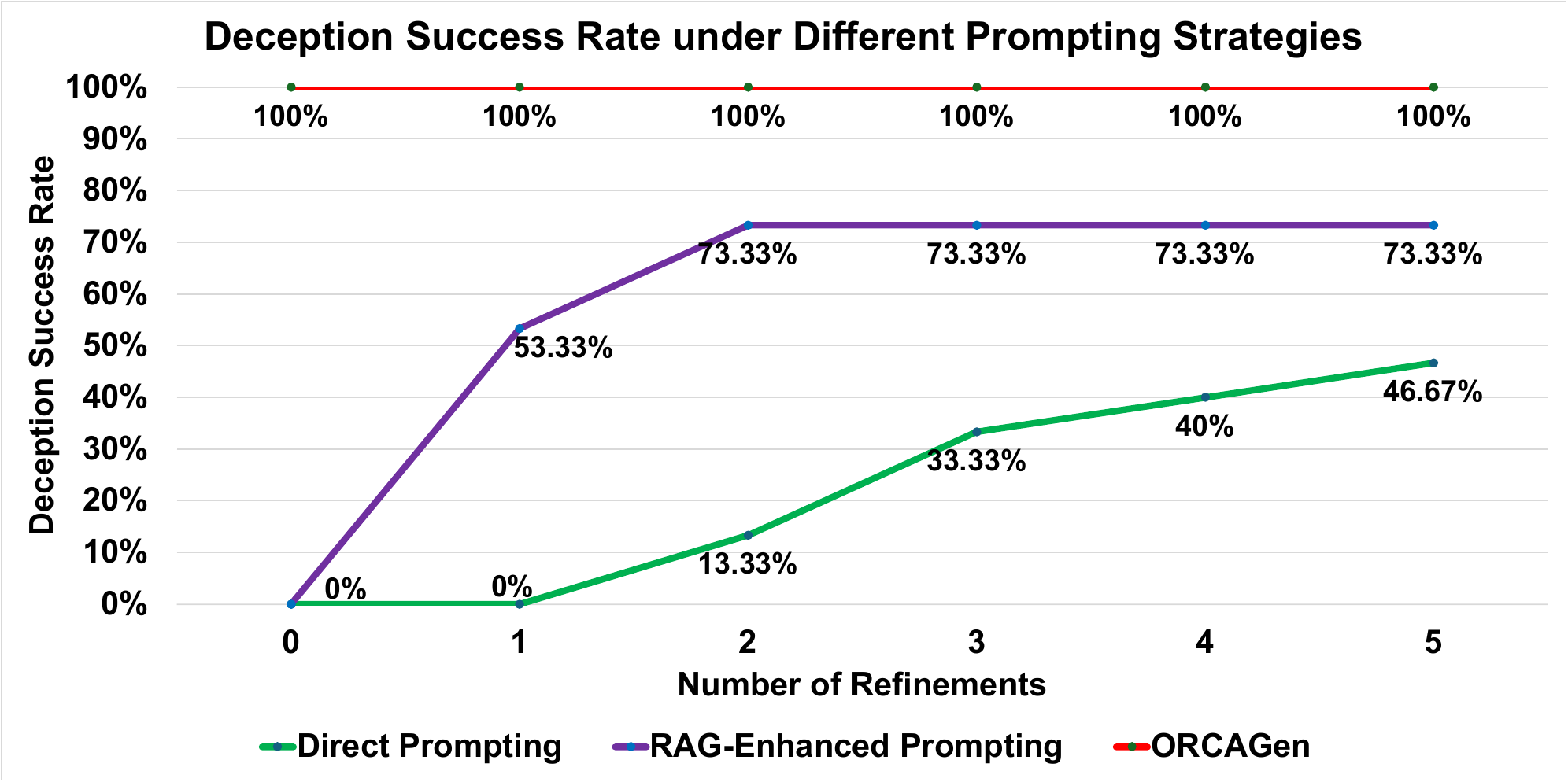}
    \vspace{-0.2cm}
    \caption{Ablation Study Comparing Different Prompting Strategies.}
    \vspace{-0.18cm}
    \label{fig:prompting_strategy_comparison}
\end{figure}

Figure~\ref{fig:prompting_strategy_comparison} shows that Direct Prompting failed in zero-shot execution and reached only 46.67\% deception success after five refinements. RAG-Enhanced Prompting improved substantially, reaching 53.33\% after one refinement and 73.33\% after two refinements, but did not improve further. ORCAGen achieved 100\% deception success in zero-shot execution. These results show that retrieval alone improves generation, but full reliability requires combining retrieved knowledge with a structured prompt and validation-oriented code scaffold.

\begin{figure}[!ht]
    \centering
    \vspace{-0.2cm}
    \includegraphics[width=0.95\columnwidth]
    {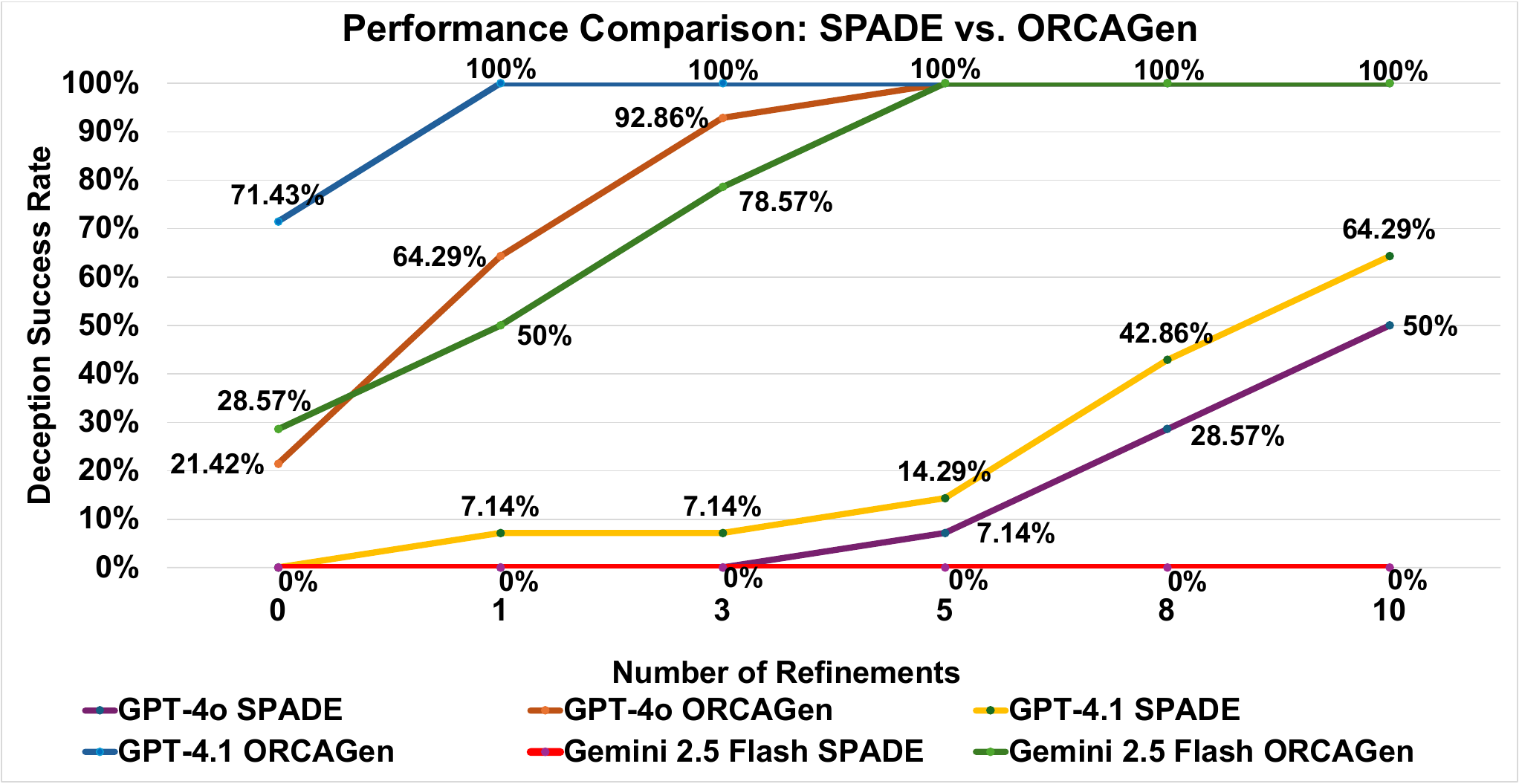}
    \vspace{-0.25cm}
    \caption{Performance Comparison between SPADE and ORCAGen.}
    \vspace{-0.2cm}
    \label{fig:spade_vs_orcagen}
\end{figure}

We also compare ORCAGen with SPADE~\cite{ahmed2025spade}, a GenAI-driven deception system that generates adaptive deception ploys using structured prompts. SPADE focuses on generating deception artifacts such as honeyfiles, patches, and API hooks, but it does not generate malware PoCs or validate generated logic against malware behavior. To preserve comparability with the prior study, we reproduced 14 representative scenarios from the SPADE evaluation and measured deception success across increasing prompt refinements, allowing a maximum of ten refinements per scenario. Figure~\ref{fig:spade_vs_orcagen} compares SPADE and ORCAGen for GPT-4o, GPT-4.1, and Gemini 2.5 Flash.

ORCAGen outperformed SPADE across all evaluated models and refinement thresholds. For GPT-4o, ORCAGen reached 64.29\% success after one refinement and 100\% after five refinements, while SPADE reached only 7.14\% after five refinements and 50\% after ten. For GPT-4.1, ORCAGen achieved 71.43\% zero-shot success and reached 100\% after one refinement, while SPADE remained below 65\% even after ten refinements. For Gemini 2.5 Flash, SPADE produced no successful deception across all refinement levels, whereas ORCAGen improved from 28.57\% zero-shot success to 100\% after five refinements. Together, these comparisons show that ORCAGen's gains come from the combination of RAG-grounded threat knowledge, structured prompt design, and validation against executable PoC behavior. Compared with direct prompting and RAG-only prompting, ORCAGen reduces refinement effort. Compared with SPADE, ORCAGen improves deployability by testing generated orchestration code against active malware behavior before adding it to the playbook.

\subsection{Performance Measurement}
This experiment measures the cost and reliability of ORCAGen's offline generation workflow and the runtime overhead of the generated orchestration logic. We evaluate four metrics across GPT-4o, GPT-5.5, Gemini 3.5 Flash, Qwen3-Coder (30B), and Claude Sonnet 4.5: \textbf{prompt refinements}, \textbf{hallucination rate}, \textbf{runtime overhead}, and \textbf{response time}. Prompt refinements measure the average number of regeneration or correction steps required before the generated PoC and orchestration code became functional. Hallucination rate measures the average occurrence of non-existent, misused, or syntactically invalid APIs, function calls, or data references. Runtime overhead measures the additional latency introduced by the validated orchestration DLL after injection into the target process. Response time measures the end-to-end latency required by each model to generate both the PoC malware and the corresponding deception orchestration code. For each scenario, ORCAGen generated a PoC malware sample and the corresponding deception orchestration code using the structured prompt and RAG context. Generated code was reviewed, compiled, and tested before being accepted. Runtime overhead was measured after injecting the validated orchestration DLL into the target process inside the virtual test environment, and each overhead value was averaged across ten independent trials.

\begin{table}[h]
\centering
\renewcommand{\arraystretch}{1.3} 
\setlength{\tabcolsep}{4pt} 
\scriptsize 
\begin{tabular}{|p{2.1cm}|p{1.8cm}|p{1.4cm}|p{1.0cm}|p{1.0cm}|}
\hline
\textbf{GenAI Model} & \textbf{Prompt Refinements (Avg.)} & \textbf{Hallucination Rate (Avg.)} & \textbf{Overhead (ms)} & \textbf{Response Time (s)} \\ \hline
GPT-4o & 0.47 & 0 & 84.35 & 71.4 \\ \hline
GPT-5.5 & 0.07 & 0 & 27.69 & 54.21 \\ \hline
Gemini 3.5 Flash & 0.27 & 0 & 18.32 & 34.92 \\ \hline
Qwen3-Coder (30B) & 1.73 & 0.27 & 183.56 & 71.32 \\ \hline
Claude Sonnet 4.5 & 0.53 & 0 & 122.68 & 74.5 \\ \hline
\end{tabular}
\vspace{1mm}
\caption{Performance of ORCAGen across different GenAI Models}
\vspace{-5mm}
\label{tab:performance_genai}
\end{table}

\textbf{Results and Discussion:}
Table~\ref{tab:performance_genai} shows clear differences in generation reliability and runtime cost across models. GPT-5.5 required the fewest refinements (0.07/task), produced no observed hallucinations, and maintained low overhead (27.69ms) with a 54.21s response time. Gemini 3.5 Flash achieved the fastest response time (34.92s) and lowest overhead (18.32ms), while requiring only 0.27 refinements and producing no hallucinations. GPT-4o was reliable but slower, with 0.47 refinements, no hallucinations, and 84.35ms overhead. Claude Sonnet 4.5 also produced no hallucinations and required 0.53 refinements, but had higher overhead (122.68ms) and the slowest response time (74.5s). Qwen3-Coder showed the weakest reliability, requiring 1.73 refinements, producing a hallucination rate of 0.27, and introducing the highest overhead (183.56ms). Overall, GPT-5.5 was strongest for rapid playbook construction, Gemini 3.5 Flash was most efficient in response time and runtime overhead, and Qwen3-Coder required the most post-processing, suggesting that local models may need stronger retrieval control, stricter prompt constraints, or additional validation support for reliable system-level code generation.

%% file: IEEE-TPS/file-related-work.tex
\section{Related Work}\label{sec:rw}
Cyber deception has long been used to mislead attackers, delay intrusions, and collect intelligence without exposing production assets. Early techniques such as honeypots, honeytokens, and honeyfiles created controlled artifacts or environments that attract malicious interaction and reveal attacker behavior~\cite{anagnostakis2005detecting, petrunic2015honeytokens, godakanda2024honey}. Later work improved adaptability through moving target defense~\cite{pagnotta2023dolos}, game-theoretic planning~\cite{schlenker2018deceiving}, fake document generation~\cite{controllable_fake_document}, and machine-learning-based honeytoken deployment~\cite{generation_deployment_honeytoken}. These systems reduce manual effort, but many still focus on static decoys or deception selection rather than generating executable, malware-specific intervention logic.

Recent GenAI-based deception systems use LLMs to generate realistic shell responses~\cite{llm_in_shell}, interactive honeypot behavior~\cite{chatbots_honeypot_world, engaging_attackers_highly, llm_interactive_system}, adaptive deception ploys~\cite{ahmed2025spade}, decoy traffic~\cite{advanced_cyber_deception_framework}, simulated SSH responses~\cite{dont_stop_believin}, deceptive SQL responses~\cite{mysqlplot_llm_based_honeypot}, honeytokens~\cite{act_as_honeytoken}, and phishing countermeasures~\cite{scamming_the_scammers}. These systems show that GenAI can improve the realism and diversity of deception artifacts. However, they largely focus on artifact generation or interactive response generation, not on producing deception logic that is grounded in malware behavior, tested against executable adversarial behavior, and compiled into a deterministic runtime playbook. Recent work has also explored deception against LLM-powered attack agents, showing that planted artifacts and misleading environmental cues can disrupt autonomous penetration-testing or reconnaissance agents~\cite{ayzenshteyn2025cloak}. ORCAGen differs by using LLMs as an offline defender-side tool to generate and validate deception playbooks against malware behavior, while runtime enforcement remains deterministic.

RAG provides a way to ground LLM outputs in external knowledge during inference. In cybersecurity, RAG can inject threat intelligence, malware procedures, API usage patterns, and defensive strategies into the prompt context. DecoyPot~\cite{sezgin2025decoypot} combines LLMs and RAG to generate realistic API honeypot responses, while other work uses RAG to retrieve contextual examples for deception detection~\cite{boumber2024llms}. These efforts demonstrate the value of retrieval grounding, but RAG remains underexplored for active deception orchestration against malware behavior. \textbf{ORCAGen} differs from prior work by generating malware-specific deception orchestration code using RAG and structured prompting. The generated code is validated offline against behavior-specific PoCs before being compiled into a runtime playbook. This separates GenAI-assisted construction from deterministic runtime enforcement, allowing ORCAGen to benefit from LLM generation without relying on live LLM inference during malware execution.



%% file: IEEE-TPS/file-conclusion.tex
\section{Discussion and Conclusion}\label{sec:conclusion}

This paper presented ORCAGen, a GenAI-assisted framework for constructing malware-specific deception playbooks. ORCAGen generates malware-specific deception orchestration code, validates it against behavior-specific PoCs, and adds it to a runtime playbook, which enforces only pre-validated deception logic without live LLM inference. To achieve this, ORCAGen uses structured knowledge bases, RAG, and prompt engineering. The evaluation shows that RAG-grounded structured prompting improves deception code generation. GPT-5.5 required the fewest refinements and produced no observed hallucinated APIs, while Gemini 3.5 Flash achieved the fastest response time and lowest runtime overhead. The ablation study shows that ORCAGen outperforms direct and RAG-enhanced prompting, and the real-world malware evaluation suggests that validated playbooks can be transferred to samples that interact with monitored credential, file-system, or API-level targets.

ORCAGen has a few limitations. Its output quality depends on KB coverage and model reliability, and the current prototype focuses on user-space malware rather than kernel-level, firmware-level, or hardware-assisted threats. The evaluation uses controlled virtual environments, which provide safety and repeatability but do not fully capture adversarial adaptation or enterprise deployment complexity. Future work will expand the KB with additional malicious behaviors such as API obfuscation, dynamic resolution, staged execution, anti-hooking, and related implementation patterns, so that generated PoCs better reflect real-world malware. We also plan to unify the external RAG layer to ensure consistent information retrieval across cloud-hosted and local models.



